\documentclass{ws-ijmpcs}

\def\trimmarks{}

\newcommand{\be}{\begin{equation}}
\newcommand{\ee}{\end{equation}}
\newcommand{\bea}{\begin{eqnarray}}
\newcommand{\eea}{\end{eqnarray}}

\def\lsim{\mathrel{\rlap{\lower4pt\hbox{\hskip1pt$\sim$}}\raise1pt\hbox{$<$}}}
\def\gsim{\mathrel{\rlap{\lower4pt\hbox{\hskip1pt$\sim$}}\raise1pt\hbox{$>$}}}
\def\nostrocostruttino#1\over#2{\mathrel{\mathop{\kern 0pt \rlap
{\hbox{$#1$}}} \hbox{\kern-.135em $#2$}}}

\newcommand{\mup}{\kappa}

\begin{document}

\markboth{M. Boglione \& J.O. Gonzalez \& S. Melis  \& A. Prokudin}
{Phenomenological implementations of TMD evolution}

\catchline{}{}{}{}{}
%
%%%%%%%%%%%%%%%%%%%%%%%%%%%%%%%%%%%%%%%%%%%%%%%%%%%%%%%%%%%%%%%%%%%%

\title{Phenomenological implementations of TMD evolution}

\author{Mariaelena Boglione\footnote{Speaker}}

\address{Physics Department, University of Turin,\\INFN, Sezione di Torino,\\via P. Giuria 1, 10125 Torino, Italy\\
%City, State ZIP/Zone, Country\footnote{State completely without abbreviations, the
%affiliation and mailing address, including country. Typeset in 8 pt italic.}
boglione@to.infn.it}

\author{J. Osvaldo Gonzalez Hernandez}

\address{INFN, Sezione di Torino,\\ Via P. Giuria 1, I-10125 Torino, Italy\\	
joseosvaldo.gonzalez@to.infn.it}

\author{Stefano Melis}

\address{Physics Department, University of Turin,\\via P. Giuria 1, 10125 Torino, Italy\\melis@to.infn.it}

\author{Alexei Prokudin}

\address{Jefferson Lab, \\12000 Jefferson Avenue, Newport News, Virginia 23606, USA\\
prokudin@jlab.org}

\maketitle

%\begin{history}
%\received{Day Month Year}
%\revised{Day Month Year}
%\published{Day Month Year}
%\end{history}

\begin{abstract}
Although the theoretical set-up of TMD evolution appears to be well established, its phenomenological 
implementations still require special attention, particularly as far as the interplay between 
perturbative and non-perturbative contributions is concerned.
These issues have been extensively studied in Drell-Yan processes, where they seem to be 
reasonably under control. Instead, applying the same prescriptions and methodologies to Semi-Inclusive 
Deep Inelastic (SIDIS)  processes is, at present, far from obvious. 
Some of the controversies related to the applications of TMD Evolution to SIDIS processes 
will be discussed with practical examples, exploring different kinematical 
configurations of SIDIS experiments. 
\keywords{evolution; transverse momentum; matching.}
\end{abstract}

\ccode{PACS numbers:13.88.+e, 12.38.Bx, 13.85.Ni}

\section{Introduction}	

Calculating the cross section which describes a hadronic process %, think of Drell-Yan scattering for example, 
over a wide range of transverse momenta, $q_T$, is a highly non-trivial task. 
While perturbative QCD computations allow us to predict its behavior in the large $q_T$ region, 
diverging contributions of large logarithms arising from the emission of soft and collinear
gluons need to be resummed in the range of low $q_T$, where $q_T \ll Q$. 
This can be achieved  
applying the Collins-Soper-Sterman (CSS) soft gluon resummation scheme~[\refcite{Collins:1984kg}].  
For instance, in a Drell-Yan (DY) process $h_1 h_2 \rightarrow \ell^+\ell^- X$, we have:
\be
%&&
%\frac{1}{\sigma_0} 
\frac{d\sigma}{dQ^2dy d P_T^2}=\sigma_0^{DY} \int\frac{d^2 \boldsymbol{b}_T e^{i \boldsymbol{q}_T\cdot\boldsymbol{b_T}}}{(2\pi)^{2}}
%\Bigg\{ 
%\nonumber\\&&
\sum_j\! e^2_jW_j(x_1,x_2,b_T,Q)
%\Bigg\}
+Y(x_1,x_2,q_T,Q)\label{MasterCSS}\,,
\ee
where $Y(x_1,x_2,P_T,Q)$, the so-called ``Y-term", is the part of the %collinear 
cross section regular at small $q_T$, 
while $W_j(x_1,x_2,b_T,Q)$ resummes the radiative gluon contributions, large when $q_T\rightarrow 0$. 
Resummation is usually performed in the $b_T$ space, the Fourier conjugate of transverse momentum space, where momentum 
conservation laws can be taken into account more easily.
For DY scattering processes, where CSS was first applied and extensively 
tested~[\refcite{Balazs:1997xd,Landry:2002ix,Konychev:2005iy}], the $W$ term reads:
\be
W_j(x_1,x_2,b_T,Q)= \exp\left[S_j(b_T,Q)\right]
\sum_{i,k} C_{ji}\otimes f_{i}(x_1,C_1^2/b_T^2)\,\, 
C_{\bar{j}k}\otimes f_{k}(x_2,C_1^2/b_T^2)\,,
\label{W}
\ee
where
\begin{equation}
S_j(b_T,Q)=-\int_{C_1^2/b_T^2}^{Q^2}\frac{d \mup^2}{\mup^2}\left[A_j(\alpha_s(\mup))\ln\left(\frac{Q^2}{\mup^2}\right)+B_j(\alpha_s(\mup))\right] 
\label{S}
\end{equation}
is the Sudakov form factor. $A_j$ and $B_j$ are perturbative coefficients that can be calculated in QCD, 
% \begin{eqnarray}
% A_j(\alpha(\mu))&=&\sum_{n=1}^{\infty}\left(\frac{\alpha_s}{2\pi}\right)^nA_j^{(n)}\\
% B_j(\alpha(\mu))&=&\sum_{n=1}^{\infty}\left(\frac{\alpha_s}{2\pi}\right)^nB_j^{(n)}\,,%\\
% %C_{ji}(z,\alpha(\mu))&=&\delta_{ij}\delta(1-z)+\sum_{n=1}^{\infty}\left(\frac{\alpha_s}{2\pi}\right)^n C_{ij}^{(n)}(z)\\
% \end{eqnarray}
while $C_1=2\exp(-\gamma_E)$ and $\gamma_E$ is the Euler's constant. The subscript $j$ indicates that the coefficients are different for $q\bar{q}$ initiated
processes (like ordinary Drell-Yan) or $gg$ fusion processes (like Higgs bosons production). 
The symbol $\otimes$ in Eq.~(\ref{W}) represents the usual collinear convolution of the Wilson coefficients $C_{ji}$ (calculable in QCD) and the collinear PDFs $f_i(x,C_1/b_T)$. 
For more details on soft gluon resummation for Drell-Yan processes see, for example, Ref.~[\refcite{Kawamura:2007gh}].

\section{Resummation in Semi-Inclusive Deep Inelastic Scattering}

For Semi-Inclusive Deep Inelastic Scattering (SIDIS) processes, $\ell N \rightarrow \ell h X$, a similar CSS expression holds 
\be
\frac{d\sigma}{d x  d z d Q^2 d^2 q_{T}} = \sigma_0^{S\!I\!D\!I\!S} \!\!\int\frac{d^2 \boldsymbol{b}_T e^{i \boldsymbol{q}_T\cdot\boldsymbol{b_T}}}{(2\pi)^{2}} 
\sum_j\! e^2_jW_j^{S\!I\!D\!I\!S}(x,z,b_T,Q)+Y^{S\!I\!D\!I\!S}%(x,z,P_T,Q)\!
\,,\label{SIDIS-CSS}
%\\
%\!\!\!\!&=&\!\!\!\!\sigma_0 \Bigg\{\!\!\int\frac{d^2 \boldsymbol{b}_T e^{i \boldsymbol{q}_T\cdot\boldsymbol{b_T}}}{(2\pi)^{2}}
%\sum_j\! e^2_j W_j^{SIDIS}(x,z,b_*,Q)F_{N\!P}(x,b_T,Q)+Y^{SIDIS}\!\Bigg\}\nonumber\\
\ee
where $q_T$ is the virtual photon momentum. Notice that, for SIDIS, we most commonly refer to the transverse momentum 
$P_T$ of the final detected hadron, $h$, in the $\gamma^* N$ c.m. frame, rather than to the virtual photon momentum $q_T$, in the $Nh$ c.m. frame. 
They are simply related by the hadronic lightcone momentum fraction $z$ through the expression $P_T=z \, q_T$, so that
\be
\frac{d\sigma}{d x  d z d Q^2 d^2 P_{T}}=\frac{d\sigma}{d x  d z d Q^2 d^2 q_{T}}\frac{1}{z^2}\,.
\ee
The resummed term $W_j^{SIDIS}$, in complete analogy to Eq.~(\ref{W}), is defined as 
\begin{equation}
W_j(x_1,x_2,b_T,Q)= \sum_{i,k}\exp\left[S_j^{S\!I\!D\!I\!S}(b_T,Q)\right]
\,C_{ji}\otimes f_{i}(x,C_1^2/b_T^2)
\,C_{kj}%^{out}
\otimes D_{k}(z,C_1^2/b_T^2),
\end{equation}	
and $D_{k}(z)$ represent the collinear unintegrated fragmentation functions (FF).
%Notice:
%$C_{ji}=C_{j\leftarrow i}$, $C_{kj}^{(out)}=C_{k\leftarrow j}^{(out)}$

Indeed, the resummed term of the cross section, $W$, cannot describe the whole $P_T$ range: 
it sums all known logarithmic terms dominating the low $P_T$ region, but does not take into account
the full fixed order, Next to Leading Order (NLO) corrections, which are important at large $P_T$ values 
(notice that here NLO means first order in $\alpha_s$ of the collinear QCD cross section). 
Because of the oscillatory nature of the Fourier integrand in Eq.~(\ref{W}), $W$ may (and does, as we shall see) become 
negative, i.e. unphysical, at large $P_T$ values. Therefore,  for a consistent description of the scattering process over the 
whole $P_T$ range, we need to {\it match} the resummed cross section with the NLO (fixed order) cross section. 
Unfortunately, however, there is no unique and indisputable matching prescription. On the contrary, as we will illustrate in what follows, 
the detailed behavior of the resummed cross section strongly depends on the process under consideration, on the energy 
at which this process takes place and on its detailed kinematics.  

To match the cross section at low and large $q_T$, the NLO cross section is usually separated into an ``asymptotic part'', 
$d\sigma^{ASY}$, which includes all the logarithmic contributions, badly diverging at small $q_T$, proportional to 
%\be 
$ \frac{Q^2}{q_T^2} \Big[A \log(\frac{Q^2}{q_T^2}) + B \Big]$, 
%\ee
and a finite part $Y$, so that
\be
\frac{d\sigma^{NLO}}{d x  d z d Q^2 d^2 q_{T}} =\frac{d\sigma^{ASY}}{d x  d z d Q^2 d^2 q_{T}} + Y\,,
\ee
and inverting
\be
Y = \frac{d\sigma^{NLO}}{d x  d z d Q^2 d^2 q_{T}} - \frac{d\sigma^{ASY}}{d x  d z d Q^2 d^2 q_{T}} \,,
\ee
see Ref.~[\refcite{Koike:2006fn}] for further details.
Now, {\it if} in the region where $P_T \simeq Q$ the resummed cross section happens to be equal or very similar 
to its asymptotic counterpart, $d\sigma^{ASY}$, then the cross section in Eq.~(\ref{SIDIS-CSS}), which we will indicate 
$W+Y$ in a simple short-hand notation, in that particular region, will almost exactly  {\it match} the NLO cross 
section, $d\sigma^{NLO}$ 
\be
W + Y \to d\sigma^{ASY} + Y = d\sigma^{ASY} + d\sigma^{NLO} - d\sigma^{ASY} = d\sigma^{NLO}\,,
\ee
and the resummed cross section can be matched to the NLO, purely-perturbative, contribution~[\refcite{Koike:2006fn}]. 
Let us stress that this matching prescription at $P_T \simeq Q$ only works if $W \sim d\sigma^{ASY}$ over a non-negligible range 
of $P_T$ values, as the matching should be {\it smooth} as well as continuous. 

\begin{figure}[t]
\centerline{
\includegraphics[width=7.5cm]{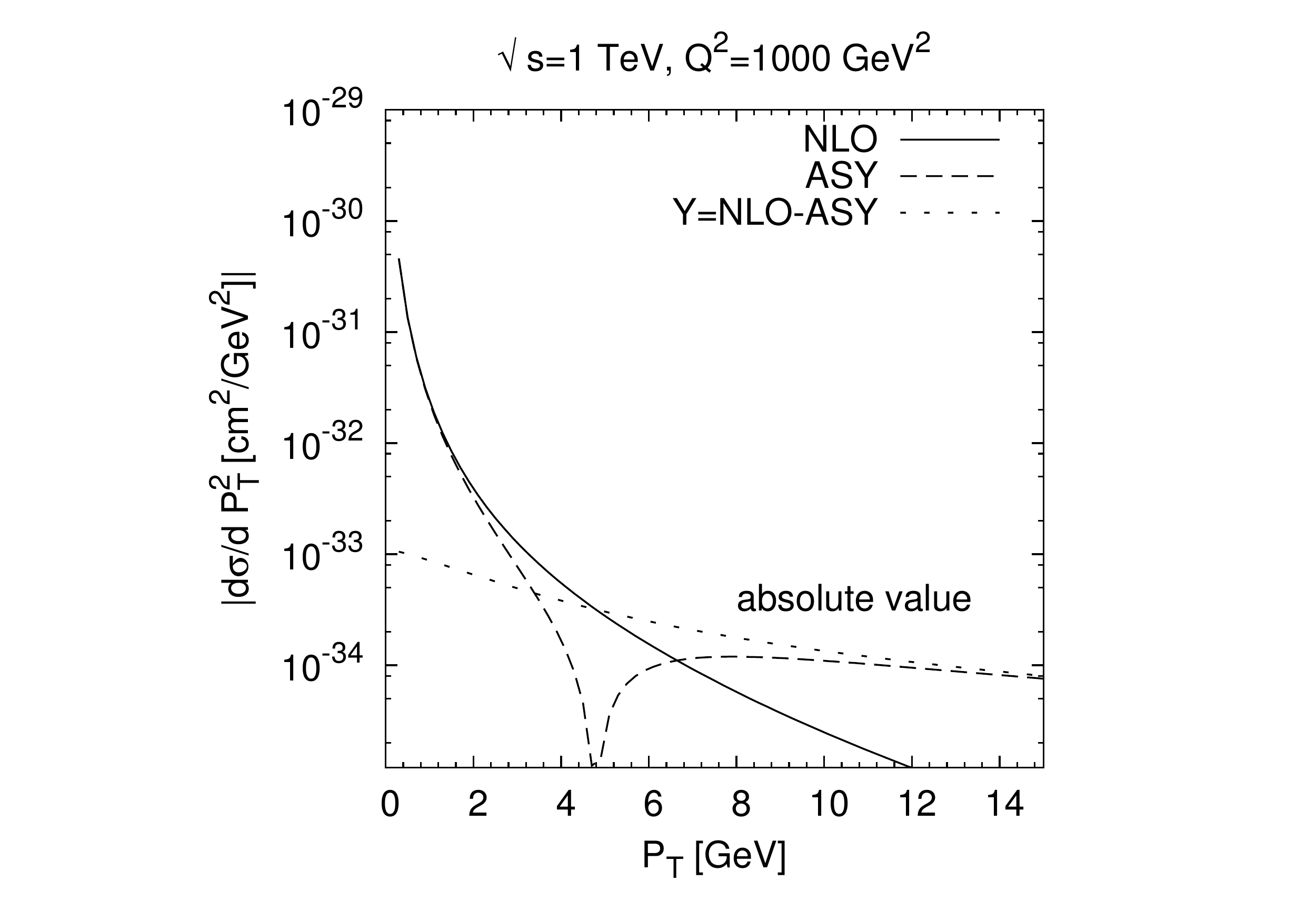} \hspace*{-1.2cm}
\includegraphics[width=7.5cm]{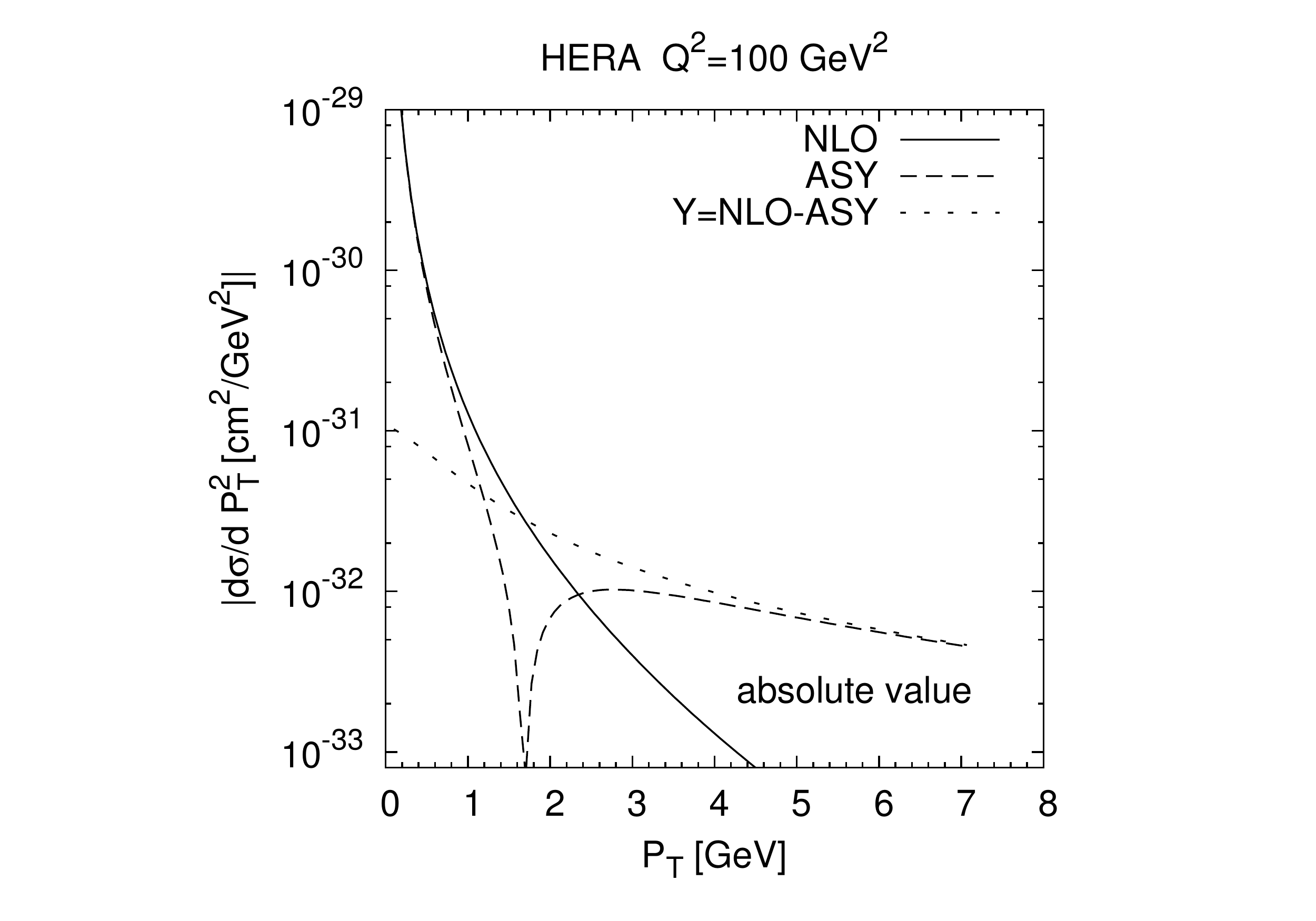}}
\vspace*{8pt}
\caption{Perturbative contributions to the SIDIS cross sections, $d\sigma^{ASY}$, $d\sigma^{NLO}$ and $Y$ factor, corresponding to two   
different SIDIS kinematical configurations: on the left panel, $\sqrt{s}=1$ TeV and $Q^2=1000$ GeV$^2$, and on the right panel  
$\sqrt{s}=300$ GeV and $Q^2=100$ GeV$^2$. Notice that, when $d\sigma^{ASY}$ becomes negative at large $P_T$, the 
$Y$ factor can become much larger than $d\sigma^{NLO}$ in that region, as $Y=d\sigma^{NLO}-d\sigma^{ASY}$. \label{f1}}
\end{figure}

Fig.~\ref{f1} shows the $d\sigma^{ASY}$, $d\sigma^{NLO}$ and Y cross section contributions for a SIDIS process in two different kinematical 
configurations: the first, on the left, at extremely high energy and large $Q^2$, $\sqrt{s}=1$ TeV and $Q^2=1000$ GeV$^2$, 
the second at a more moderate energy and $Q^2$, similar to what one could expect at a HERA-like experiment, $\sqrt{s}=300$ GeV and $Q^2=100$ GeV$^2$. 
Notice that, as $d\sigma^{ASY}$ becomes negative at large $P_T$ (on the log-plots we can only show its absolute value), 
the $Y$ term can become much larger than the $NLO$ cross section in that region, as $Y=d\sigma^{NLO}-d\sigma^{ASY}$.

At this stage one should wonder whether, given a well-defined SIDIS scattering process, a kinematical range in 
which $W \sim d\sigma^{ASY}$ actually does exist, where the matching can successfully be performed.
However, before we can answer this question we should worry about the {\it non-perturbative} contributions to 
the Sudakov factor, Eq.~(\ref{S}).
In fact, as the CSS formalism relies on a Fourier integral over $b_T$ which runs from 0 to $\infty$, see Eq.~(\ref{MasterCSS}), 
no prediction can be made without an ansatz prescription for the non-perturbative region,
where $b_T$ is large and $P_T$ is small. According to Eq.~(\ref{S}), the Sudakov factor hits the Landau pole in $\alpha_s$ at large values of 
%diverges at large 
$b_T$, therefore in the CSS 
scheme a freezing prescription is used, which prevents $b_T$ from getting any larger than some (predefined) maximum value 
$b_{max}$: 
\begin{equation}
 b^*=\frac{b_T}{\sqrt{1+b_T^2/b_{max}^2}}\,.
\end{equation}
In addition, the lower limit of integration in Eq.~(\ref{S}) is replaced by $\mu_b=C_1/b^*$. 
Then the cross section is written as
%\begin{eqnarray}
\be
\frac{d\sigma}{dQ^2dy d q_T^2}\!=\!
%\sigma_0\Bigg\{\!\!\int\frac{d^2 \boldsymbol{b}_T e^{i \boldsymbol{q}_T\cdot\boldsymbol{b_T}}}{(2\pi)^{2}} 
%\sum_j\! e^2_jW_j(x_1,x_2,b_T,Q)+Y(x_1,x_2,q_r,Q)\!\Bigg\}  \\
\sigma_0 \int\!\!\!\frac{d^2 \boldsymbol{b}_T e^{i \boldsymbol{q}_T\cdot\boldsymbol{b_T}}}{(2\pi)^{2}}
\sum_j\! e^2_j W_j(x,z,b_*,Q)F_{N\!P}(x,z,b_T,Q)+Y(x,z,q_r,Q)
\ee
%\end{eqnarray}
where $W$, the {\it perturbative} part of the Sudakov factor, is a function of $b^*$ only, while the whole non-perturbative content 
is contained in $F_{N\!P}(x,z,b_T,Q)$, the {\it non-perturbative} part of the Sudakov factor, 
which accounts for the non-perturbative behavior of the cross section at large $b_T$ (i.e. small $P_T$).
\begin{figure}[t]
\centerline{
\includegraphics[width=6.20cm]{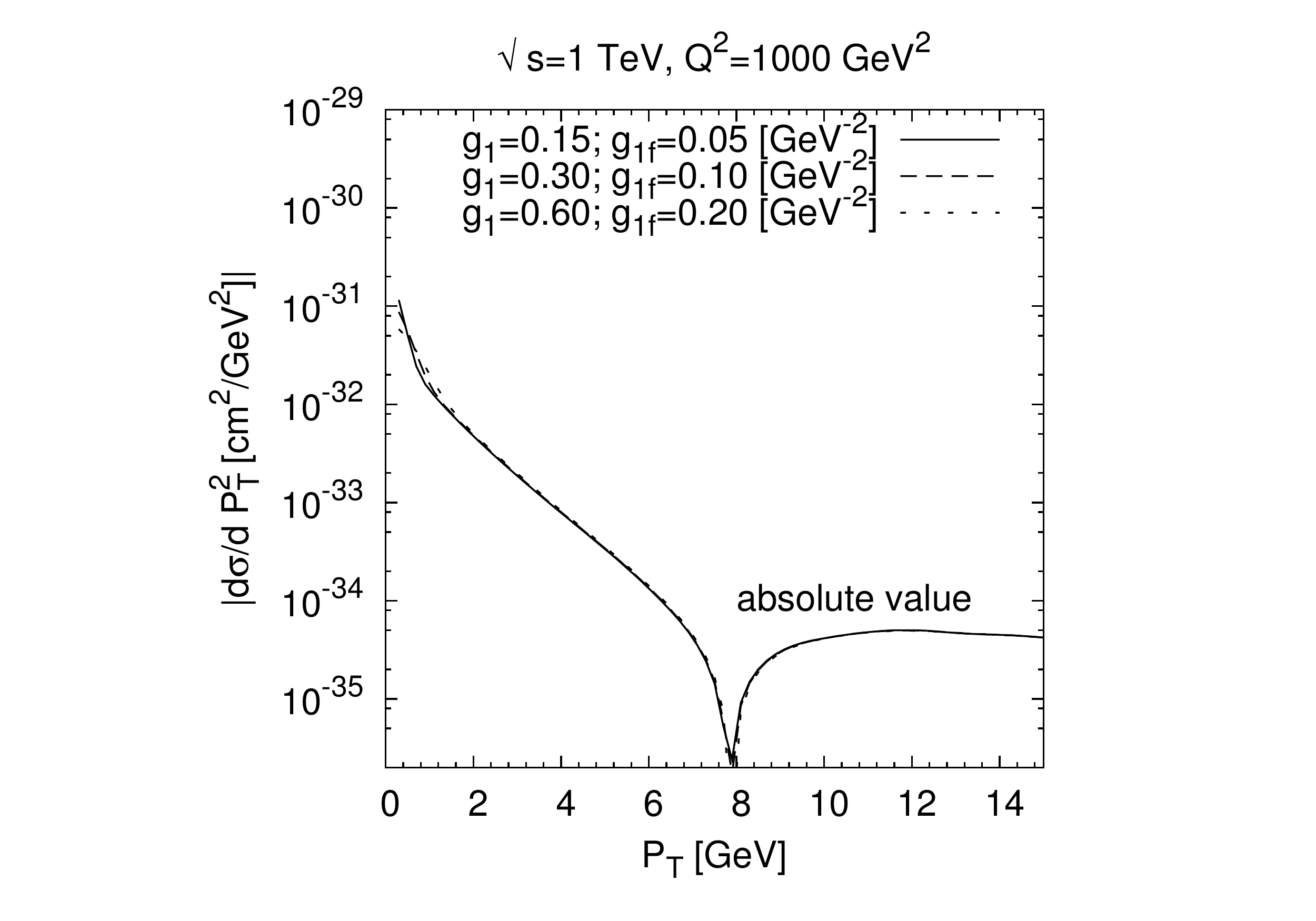} \hspace*{-2.2cm}
\includegraphics[width=6.20cm]{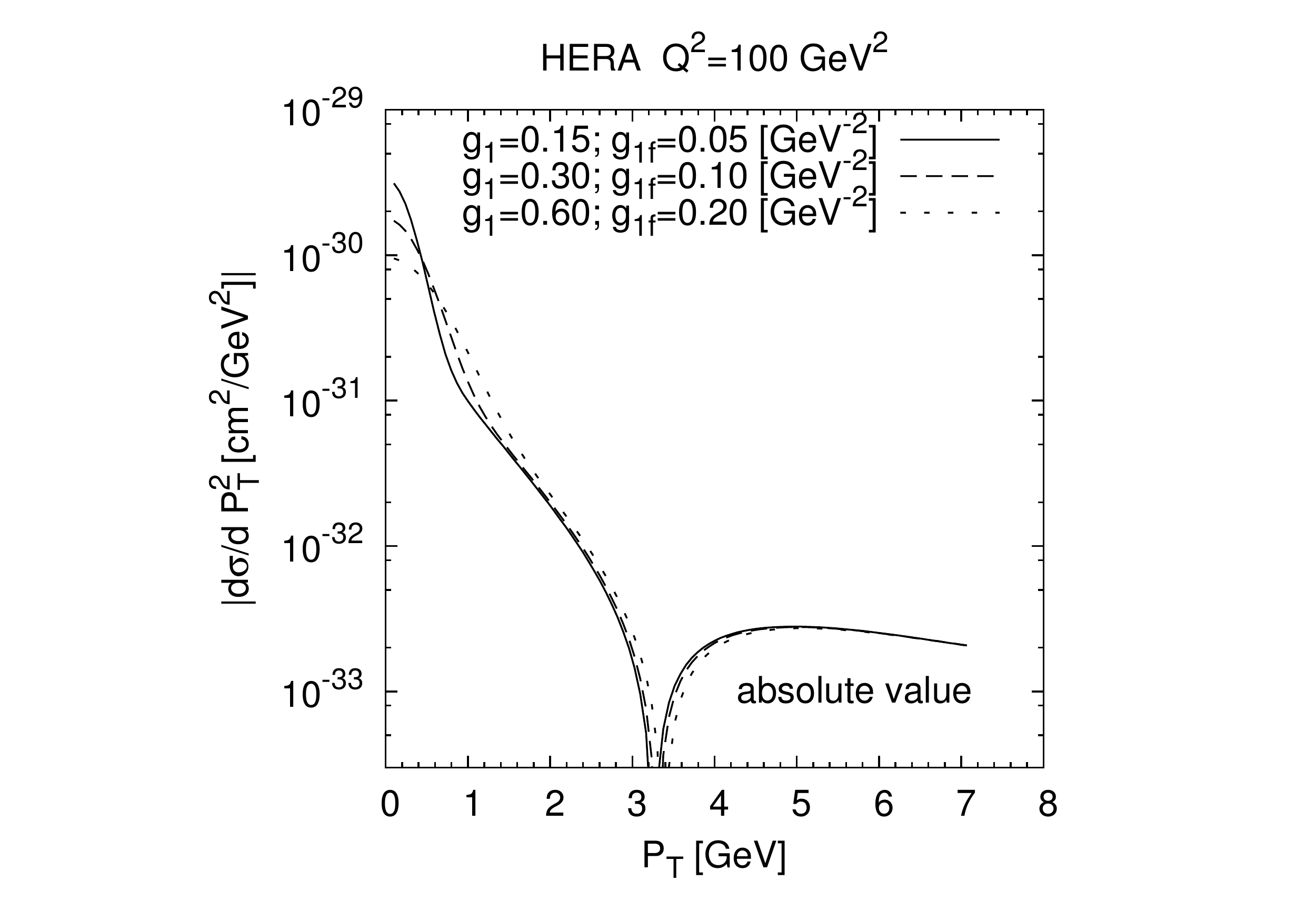} \hspace*{-2.2cm}
\includegraphics[width=6.20cm]{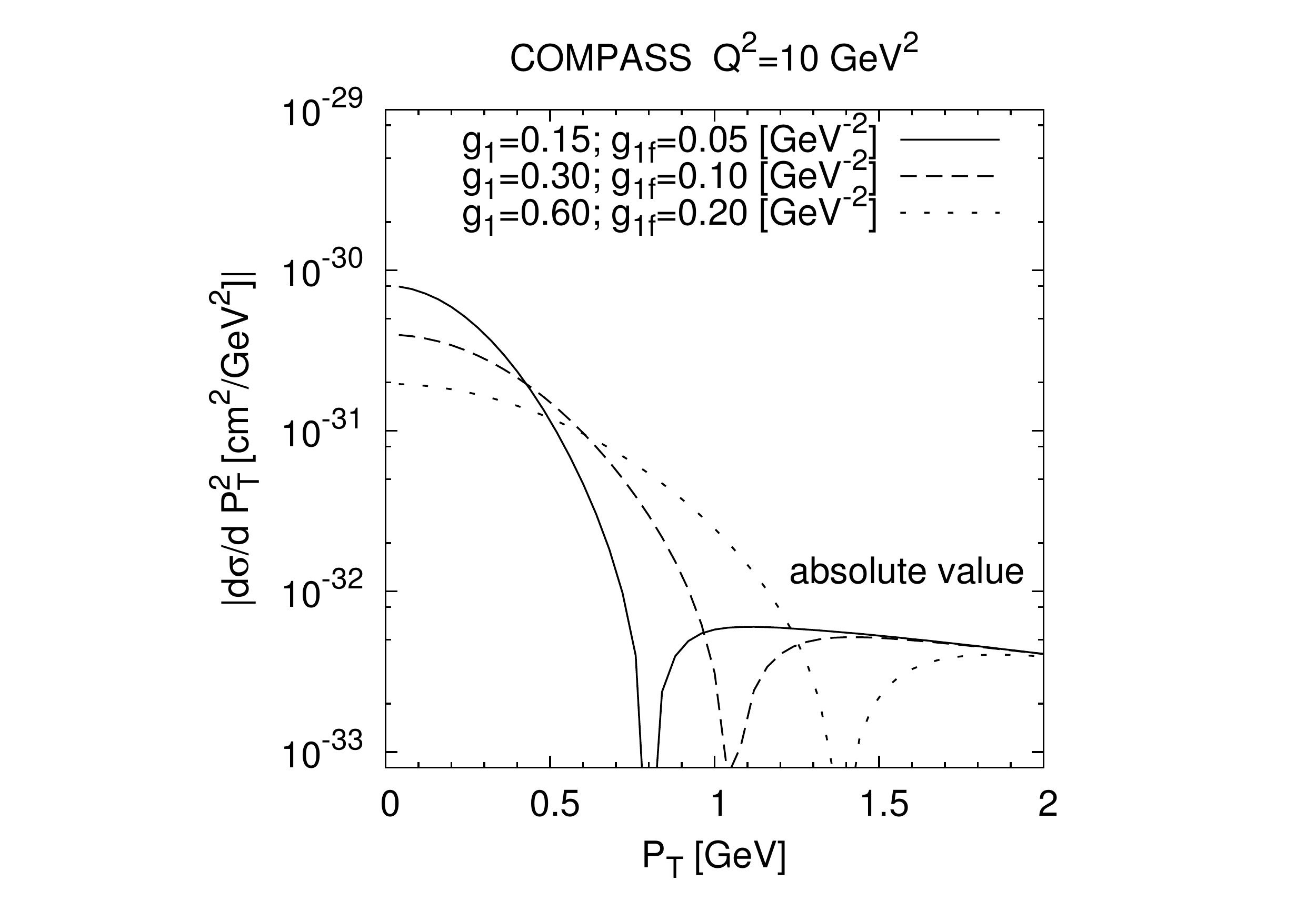}}
\vspace*{8pt}
\caption{Non-perturbative contribution, $F_{N\!P}$, to the Sudakov factor $S$, calculated at three different values of $g_1$ and $g_{1f}$, corresponding to three 
different SIDIS kinematical configurations: on the left panel, $\sqrt{s}=1$ TeV and $Q^2=1000$ GeV$^2$, in the central panel  
$\sqrt{s}=300$ GeV and $Q^2=100$ GeV$^2$, and on the right panel $\sqrt{s}=17$ GeV and $Q^2=10$ GeV$^2$. \label{f2}}
\end{figure}

As a simple illustration, let's consider a Gaussian model for the non-perturbative function $F_{N\!P}$:
\be
F_{N\!P} = \exp{\frac{1}{2} [ (- g_1 - g_{1f} / z ) b_T^2 ]}\,.
\ee
Obviously, having introduced a parametrization to represent $F_{N\!P}$, our results will now inevitably be affected 
by some degree of model dependence: how strong a model dependence is determined by the kinematics of the SIDIS process 
under consideration. Fig.~\ref{f2} shows the non perturbative contribution to the Sudakov factor, $F_{N\!P}$, 
calculated with three different values of the pair ($g_1 , g_{1f}$), and corresponding to three 
different SIDIS kinematical configurations: on the left panel, at extremely high energy and 
large $Q^2$ ($\sqrt{s}=1$ TeV and $Q^2=1000$ GeV$^2$), in the central panel  
at a more moderate energy and $Q^2$ ($\sqrt{s}=300$ GeV and $Q^2=100$ GeV$^2$) typical for example 
of an experiment like HERA, and on the right panel a low energy and $Q^2$ 
configuration, similar to the kinematics of the COMPASS experiment ($\sqrt{s}=17$ GeV and $Q^2=10$ GeV$^2$). 
These plots clearly show that, in a very large energy and $Q^2$ configuration, the non-perturbative content of the Sudakov factor, $F_{N\!P}$,
%which takes into account the non-perturbative behavior at large $b_T$ (i.e. small $P_T$), 
induces only a very mild dependence on the parameters of the model 
at small $P_T$ and the three curves change sign at the same $P_T$ value. Instead, at smaller energies and $Q^2$s, the dependence of the SIDIS cross section on the value of the model parameters becomes 
stronger and stronger, and the three curves change sign at three very different values of $P_T$. 

\begin{figure}[t]
\centerline{
\includegraphics[width=6.20cm]{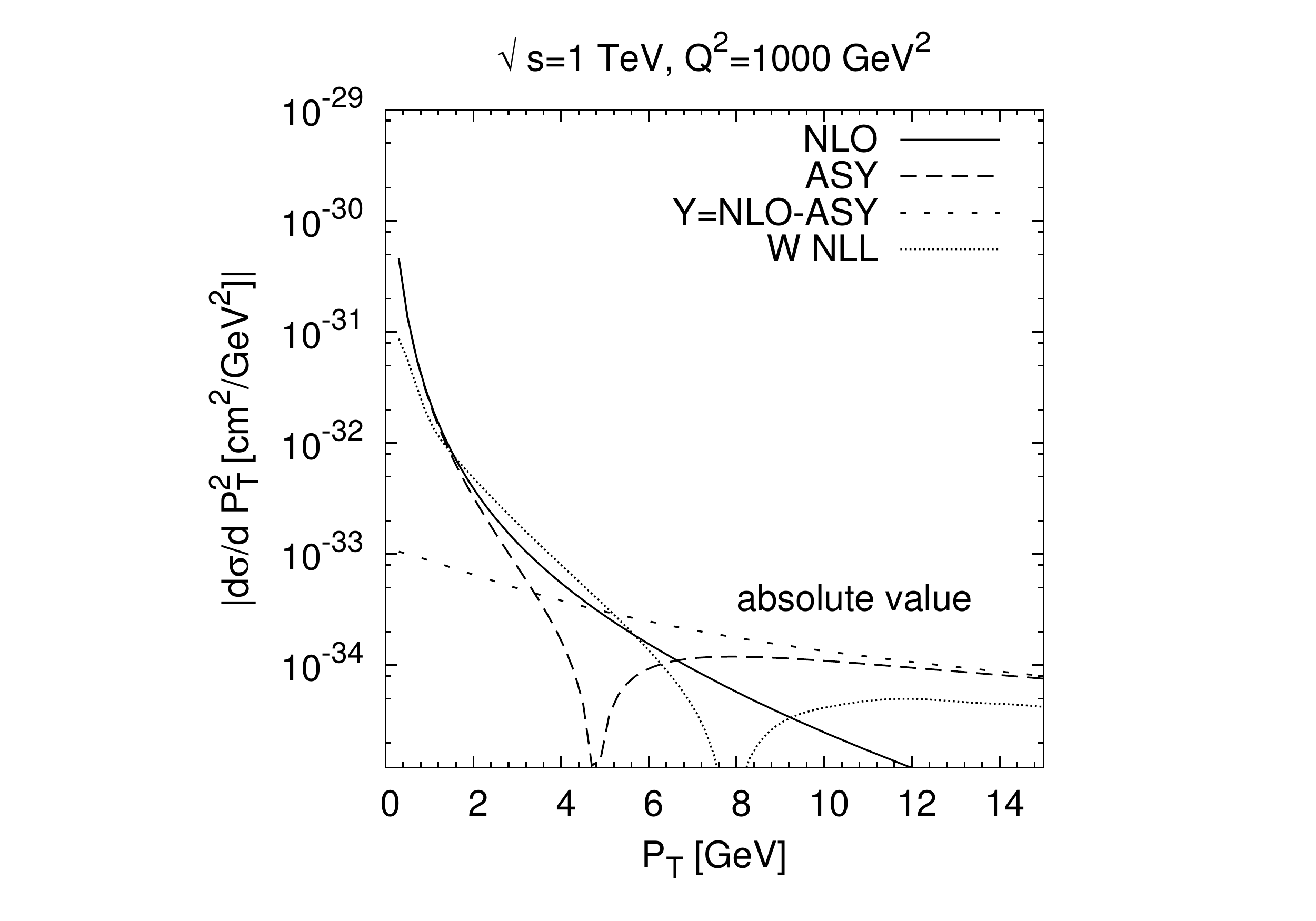} \hspace*{-2.2cm}
\includegraphics[width=6.20cm]{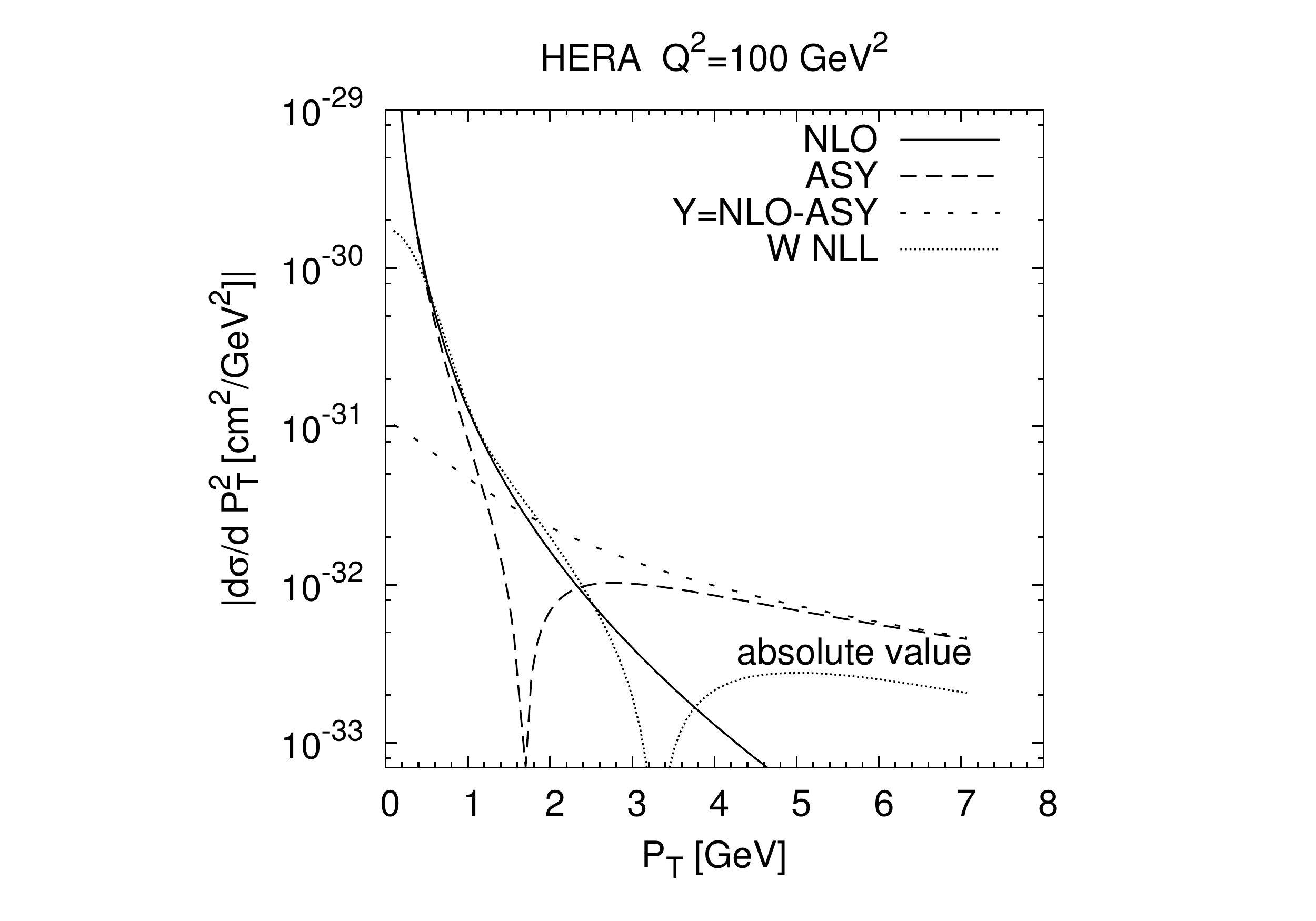} \hspace*{-2.2cm}
\includegraphics[width=6.20cm]{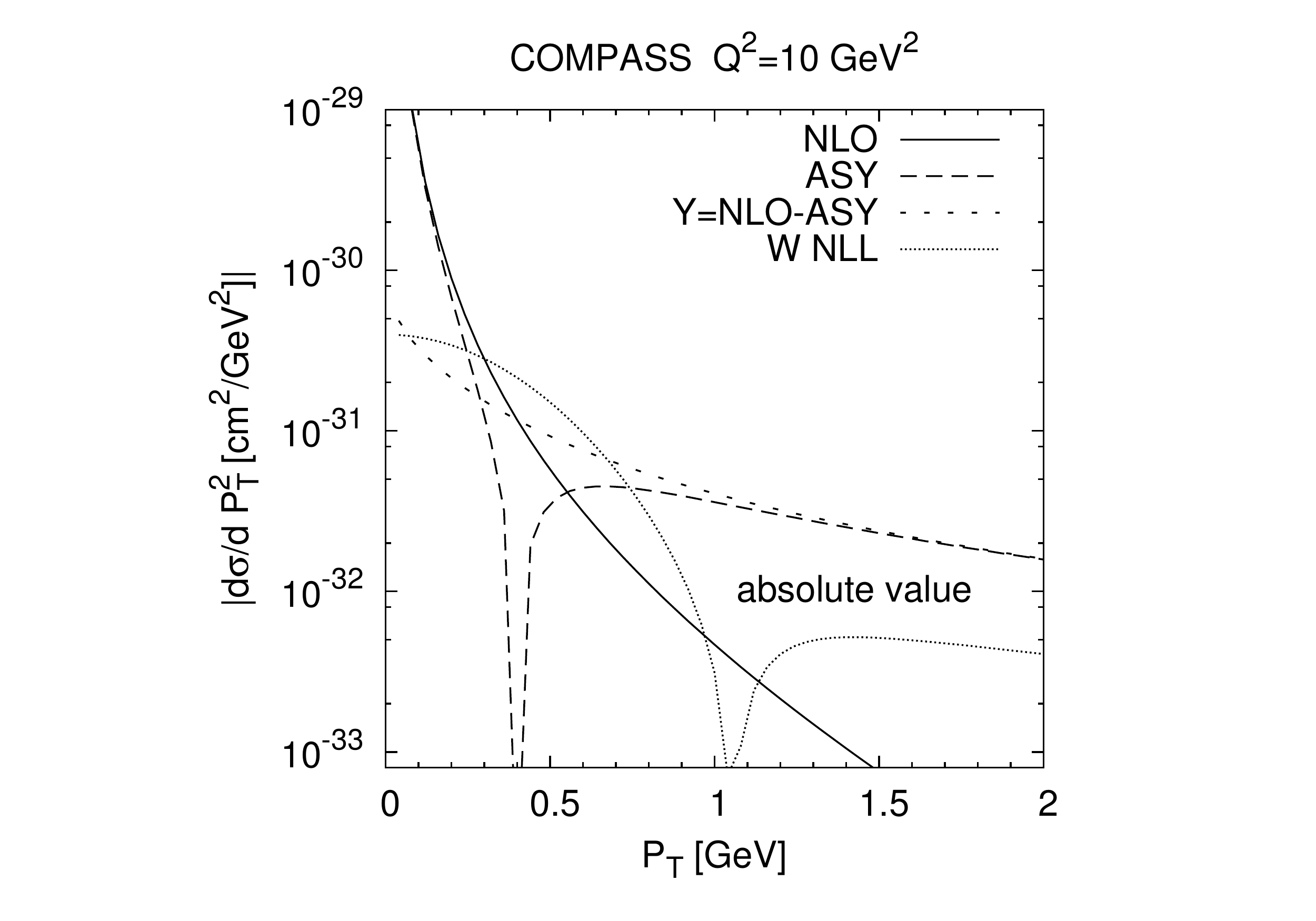}}
\vspace*{8pt}
\caption{Perturbative contributions to the SIDIS cross sections, $d\sigma^{ASY}$, $d\sigma^{NLO}$, $Y$ factor and resummed term $W^{NLL}$ corresponding to three 
different SIDIS kinematical configurations: on the left panel, $\sqrt{s}=1$ TeV and $Q^2=1000$ GeV$^2$, 
in the central panel $\sqrt{s}=300$ GeV and $Q^2=100$ GeV$^2$,
and on the right panel  $\sqrt{s}=17$ GeV and $Q^2=10$ GeV$^2$. 
Notice that, around $P_T\sim Q$, in none of these configurations, the resummed term $W$ gets even close to $d\sigma^{ASY}$, while $Y$ can be very large; 
moreover, $W$ and $d\sigma^{ASY}$ change sign at very different $P_T$s. 
 \label{f3}}
\end{figure}

It should now be perfectly clear that a successful matching heavily depends on the subtle interplay between perturbative and
non-perturbative contributions to the cross section~[\refcite{Qiu:2000hf}], and that finding a kinematical range in which the resummed cross section $W$ matches its asymptotic 
counterpart $d\sigma^{ASY}$ cannot be taken for granted. On the contrary, Fig.~\ref{f3} shows that in the three SIDIS configurations considered above, 
around $P_T\sim Q$, the resummed term, $W$ never gets even close to $d\sigma^{ASY}$, while $Y$ can be very large; moreover, $W$ and $d\sigma^{ASY}$ change sign at very different $P_T$s, and these 
specific $P_T$ values are determined by the SIDIS kinematical configuration. 
This mismatch is partly due to the non-perturbative content of the cross section, which turns out to be non-negligible even at high energies and transverse momenta.
To try and solve this problem one could experiment different, more elaborate, matching prescriptions. In alternative to $d\sigma= W + Y$, for instance, we could require
\be
d\sigma = W^{NLL} - W^{FXO} + d\sigma^{NLO}\,,
\label{match2}
\ee
where $W^{FXO}$ is the next to leading log (NLL) resummed cross section approximated at first order in $\alpha_s$,
with a first order expansion of the Sudakov exponential $S$, see Eq.~(\ref{S}). 
%
% \begin{figure}[h]
% \centerline{
% \includegraphics[width=8.5cm]{matching} }
% \vspace*{8pt}
% \caption{Matching prescription, Eq.~\ref{match2}.
% \label{f5}}
% \end{figure}
%

In the absence of non-perturbative content and in the fully perturbative limit $b_T \to 0$ (and $P_T \to \infty$), one can easily show that $W^{FXO} \to d\sigma^{ASY}$ so that, in this region, 
$d\sigma = W^{NLL} - W^{FXO} + d\sigma^{NLO} \to  W^{NLL} - d\sigma^{ASY} + d\sigma^{NLO}= W^{NLL} + Y$, and we recover Eq.~(\ref{SIDIS-CSS}).
On the other hand, $W^{FXO}$ contains the same non-perturbative content we assign to $W^{NLL}$; 
consequently we might expect to find a region in which $W^{FXO} \sim W^{NLL}$, allowing to match the SIDIS cross section $d\sigma = W^{NLL} - W^{FXO} + d\sigma^{NLO}$ to the purely perturbative 
cross section $d\sigma^{NLO}$.  

%\vspace*{3mm}
\centerline{\includegraphics[width=8.5cm]{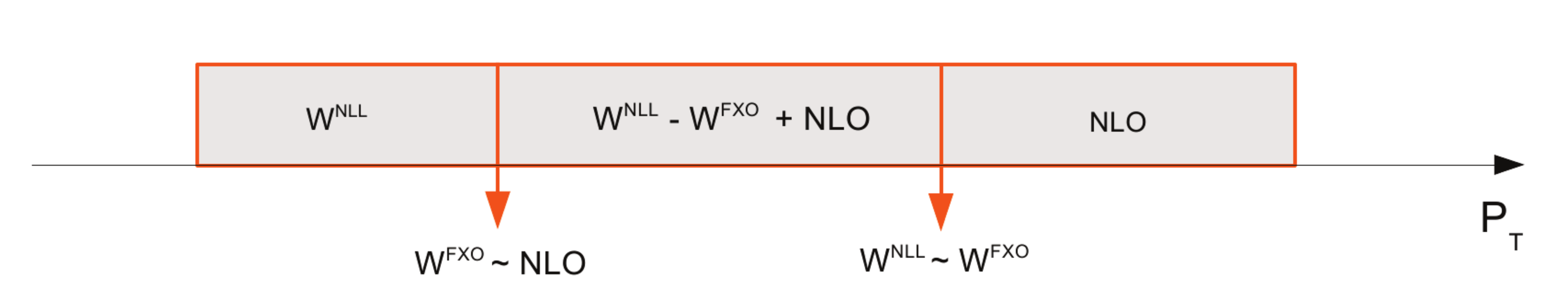}}
%\vspace*{3mm}

\begin{figure}[t]
\centerline{
\includegraphics[width=6.2cm]{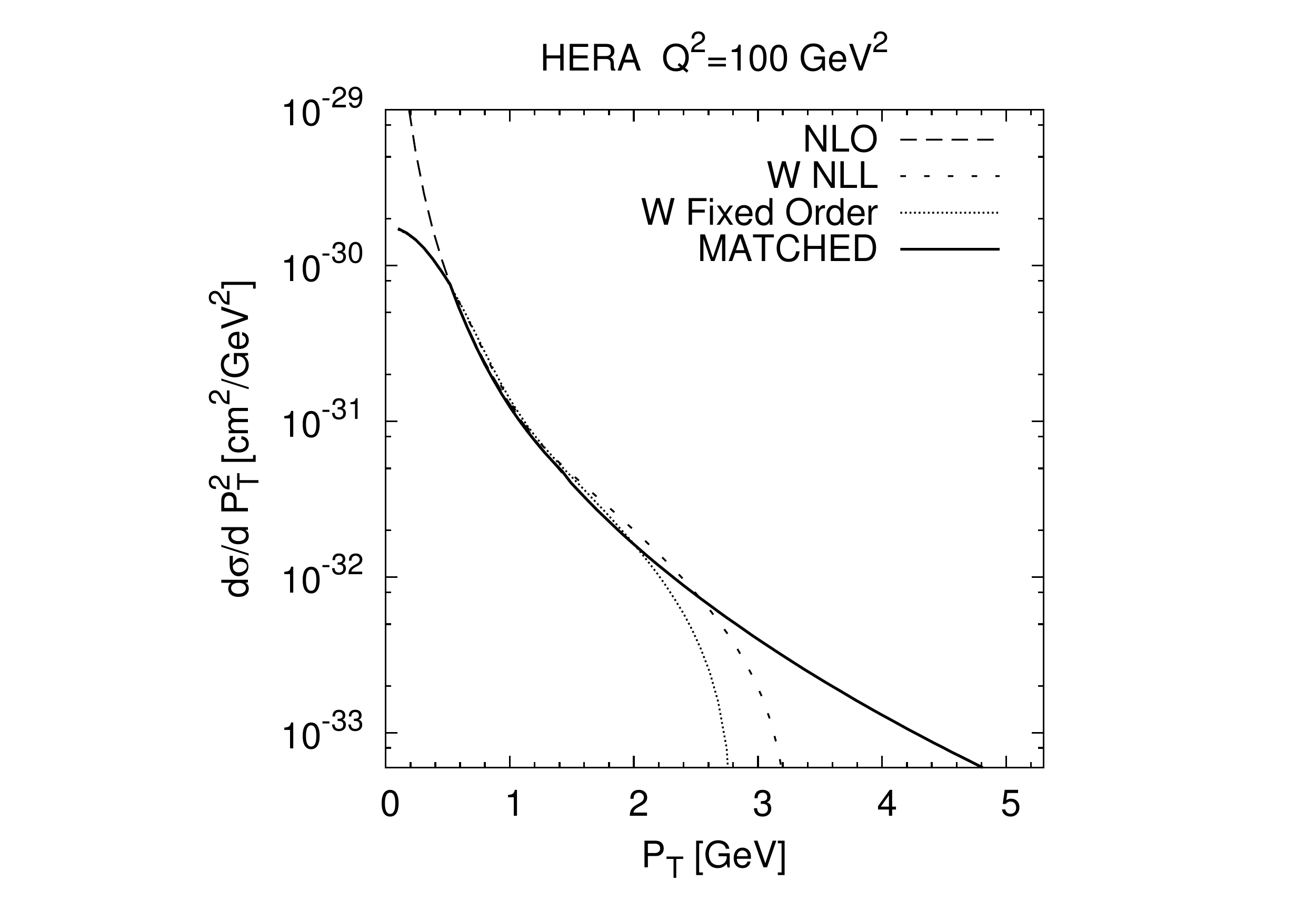}\hspace*{-1.2cm}
\includegraphics[width=6.2cm]{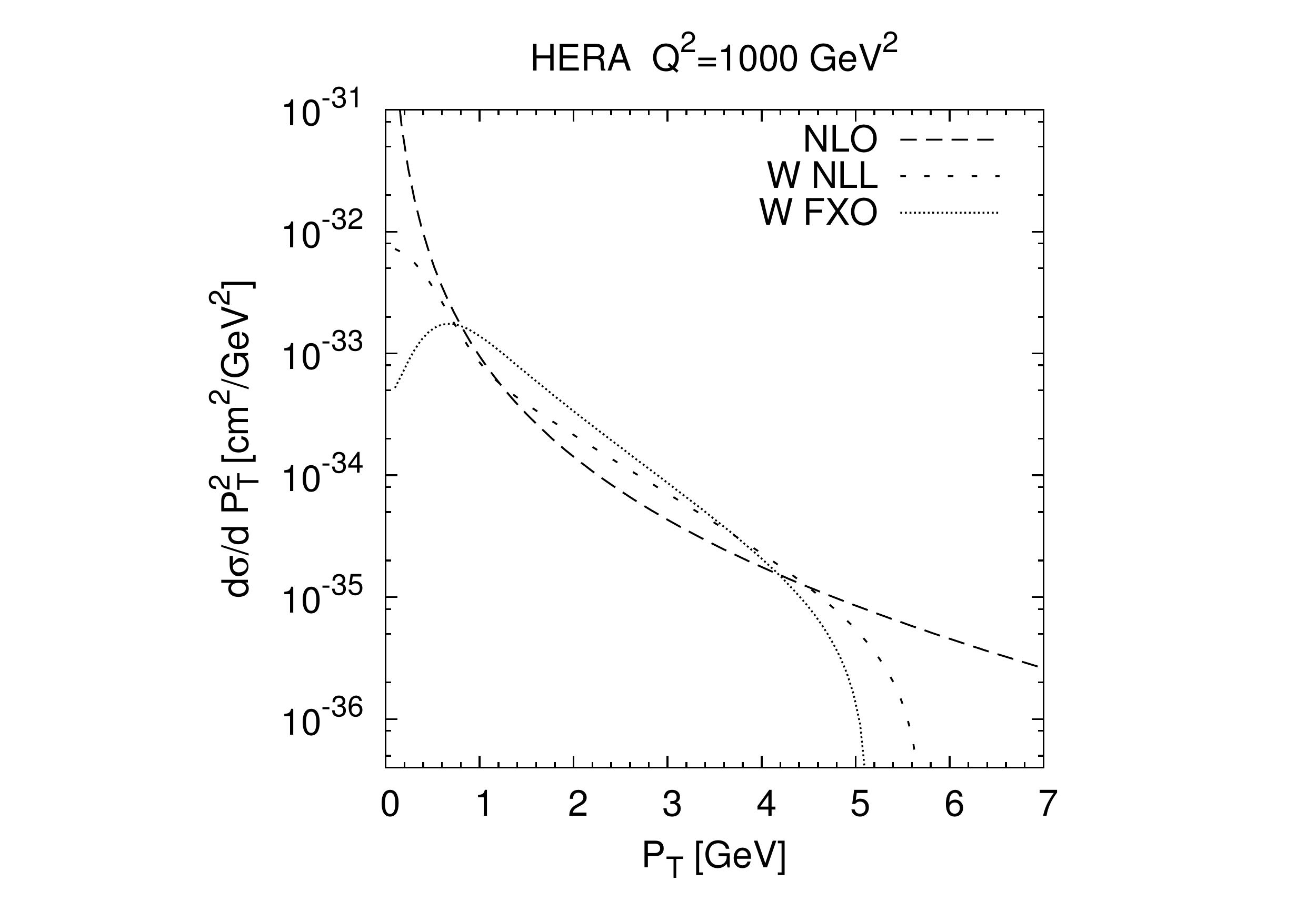}}
\vspace*{8pt}
\caption{In the HERA-like SIDIS kinematical configuration (left panel), $\sqrt{s}=300$ GeV and $Q^2=100$ GeV$^2$, 
the matching prescription of Eq.~(\ref{match2}) works remarkably well. 
Contrary to what one could naively expect, the quality of matching turns out to deteriorate at larger values of $Q^2$, 
as shown in the right panel, where $\sqrt{s}=300$ GeV and $Q^2= 1000$ GeV$^2$. 
%In the COMPASS SIDIS kinematical configuration (right panel) the matching scheme of Eq.~(\ref{match2}) is completely unsuccessful: 
%any attempt to match the cross section would result in very pronounced cusps.
\label{f4}}
\end{figure}

The left panel of Fig.~\ref{f4} shows a SIDIS configuration in which this prescription actually works remarkably well. 
At “large” $P_T$s there is a region where $W^{NLO}$ and $W^{FXO}$ are roughly the same over a
range wide enough to allow for a safe matching: %see Fig.~\ref{f5}, 
here all the curves are
reasonably close to each other and they have roughly the same curvature, allowing the matching to be smooth as well as continuous.
In addition, at  “low” $P_T$s, there is a region where $W^{NLL} \sim W^{FXO} \sim d\sigma^{NLO}$, which makes the description of this SIDIS cross section perfectly matched over the entire $P_T$ range. 
%
% 
% 
% Here in fact, at  “low” $P_T$s, there is
% a region where $W^{NLL}$ and $W^{FXO}$ are roughly the same over a
% range wide enough to allow for a safe matching: %see Fig.~\ref{f5}, 
% here all the curves are
% reasonably close to each other and they have roughly the same curvature, allowing the matching to be smooth as well as continuous. In addition, at “large” $P_T$s there
% is a region where $W^{FXO} \sim d\sigma^{NLO}$, which makes the description of this SIDIS cross section perfectly matched over the entire $P_T$ range. 
%
In light of these results, one could think that this matching procedure gets more and more successful with growing $Q^2$. 
This, unfortunately, does not seem to be the case, as shown in the right panel of Fig.~\ref{f4} for $\sqrt{s}=300$ GeV and $Q^2=1000$ GeV$^2$.
Notice that this happens also for DY processes at Tevatron kinematics, as discussed in Ref.~[\refcite{Arnold:1990yk}].

\begin{figure}[t]
\centerline{
\includegraphics[width=6.7cm]{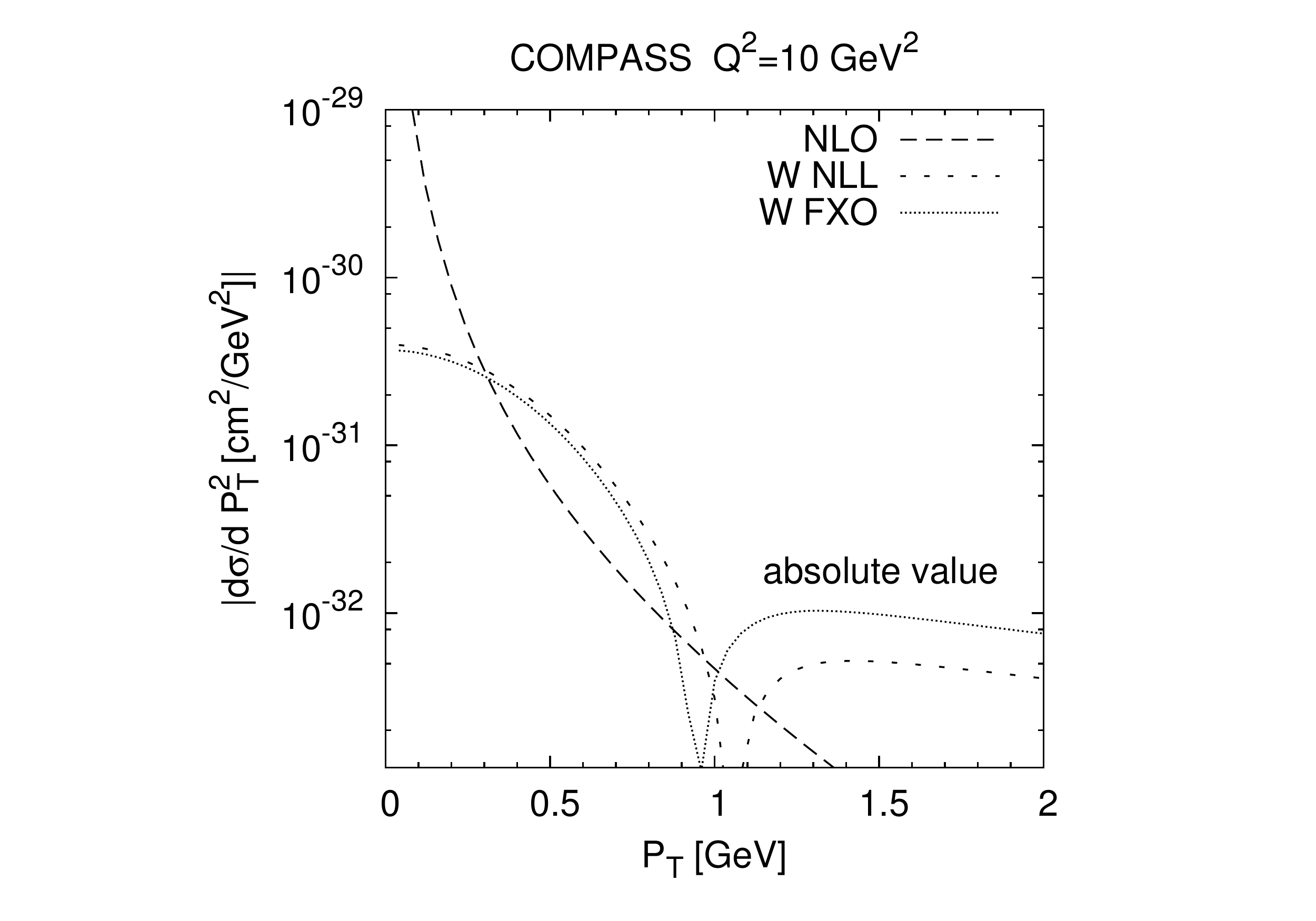}}
\vspace*{8pt}
\caption{In the COMPASS SIDIS kinematical configurations %with $\sqrt{s}=17$ GeV and $Q^2=10$ GeV$^2$, 
the matching prescription of Eq.~(\ref{match2}) does not work, as the whole cross section is dominated by its non-perturbative content.
\label{f5}}
\end{figure}

Last, but most importantly, Fig.~\ref{f5} shows what happens when the SIDIS kinematics corresponds to low energy and 
momentum transfer ($\sqrt{s}=17$ GeV and $Q^2=10$ GeV$^2$), 
as it is the case for the COMPASS experiment, where the non-perturbative regime basically dominates the whole cross section. 
Here the curves are far from each other and they have different curvatures: 
there is no way to realize a smooth matching, avoiding the appearance of pronounced ``cusps''.

\section{Conclusions}

Resummation in the impact parameter $b_T$ space is a very powerful tool. However, its successful
implementation is affected by a number of practical difficulties: the strong influence of the kinematical details of the SIDIS process, 
the possible dependence of the parameters used to model the non-perturbative content of the SIDIS cross section, 
the complications introduced by having to perform phenomenological studies in the $b_T$ space, where we loose any direct 
connection of our inputs to the exact outcomes in the conjugate $P_T$ space, etc...

Moreover, it is often very hard to define the exact boundaries of the four regions of interest:
$P_T \sim \Lambda_{QCD} \ll Q$, ~ $\Lambda_{QCD} \ll P_T \ll Q$,  ~  $P_T \sim Q$,  and  $P_T > Q$.

Indeed, matching prescriptions have to be applied to achieve a reliable description of the SIDIS process
over the full $P_T$  range, going smoothly from one region to the following.
However, the procedures analyzed in this preliminary study seem to be successful only in those cases 
where $W^{NLL}$, $W^{FXO}$, $d\sigma^{ASY}$ and $d\sigma^{NLO}$ are reasonably close to each other, as well as 
having similar curvatures, over sufficiently wide regions to allow us to switch smoothly from one to the other. 
Clearly, this can only happen when the effect of the non-perturbative contributions to the Sudakov factor, 
$F_{N\!P}$, is limited and does not stretch to the large $P_T$ region.

While for SIDIS processes at high energies and $Q^2$s, the matching prescriptions described above may or may not work, 
depending on the details of the specific kinematics under consideration, for COMPASS and HERMES data, at our present knowledge, 
these procedures can certainly not be applied without substantial refinements and adjustments.

%\bibliographystyle{ws-ijmpcs.bst}

%\bibliography{sample}

\end{document}